\documentclass[journal]{IEEEtran}
\usepackage{cite}
\usepackage{booktabs}
\usepackage{multirow}
\usepackage{adjustbox}
\usepackage{algorithm}
\usepackage{algpseudocode}
\usepackage[backref]{hyperref}
\usepackage[table, dvipsnames]{xcolor}
\usepackage{amsmath, amssymb, amsfonts, bm}
\DeclareMathOperator*{\argmin}{argmin}

\begin{document}
\title{Deep Inverse Design for High-Level Synthesis}
\author{
\IEEEauthorblockN{Ping Chang\IEEEauthorrefmark{1}, \and Tosiron Adegbija\IEEEauthorrefmark{1}, \and Yuchao Liao\IEEEauthorrefmark{1}, \and Claudio Talarico\IEEEauthorrefmark{4}, \and Ao Li\IEEEauthorrefmark{1}\IEEEauthorrefmark{2}, \and Janet Roveda\IEEEauthorrefmark{1}\IEEEauthorrefmark{2}\IEEEauthorrefmark{3}}

\IEEEauthorblockA{\IEEEauthorrefmark{1}Department of Electrical \& Computer Engineering, The University of Arizona, Tucson, AZ, USA}\\
\IEEEauthorblockA{\IEEEauthorrefmark{2}Bio5 Institute, The University of Arizona, Tucson, AZ, USA}\\
\IEEEauthorblockA{\IEEEauthorrefmark{3}Department of Biomedical Engineering, The University of Arizona, Tucson, AZ, USA}\\
\IEEEauthorblockA{\IEEEauthorrefmark{4}School of Engineering \& Applied Science, Gonzaga University, Spokane, WA, USA}
}
\maketitle

\begin{abstract}
High-level synthesis (HLS) has significantly advanced the automation of digital circuit design, yet the need for expertise and time in pragma tuning remains challenging. Existing solutions for the design space exploration (DSE) adopt either heuristic methods, lacking essential information for further optimization potential, or predictive models, missing sufficient generalization due to the time-consuming nature of HLS and the exponential growth of the design space. To address these challenges, we propose \textit{Deep Inverse Design for HLS (DID$4$HLS)}, a novel approach that integrates graph neural networks and generative models. DID4HLS iteratively optimizes hardware designs aimed at compute-intensive algorithms by learning conditional distributions of design features from post-HLS data. Compared to four state-of-the-art DSE baselines, our method achieved an average improvement of $42.8\% $ on average distance to reference set (ADRS) compared to the best-performing baselines across six benchmarks, while demonstrating high robustness and efficiency. The code is available at \url{https://github.com/PingChang818/DID4HLS}.\\

\end{abstract}

\section{Introduction}
The increasing complexity of circuit designs, compounded by stringent time-to-market imperatives, underscores the demand for advanced design automation \cite{sinha2015high}. High-level synthesis (HLS) is an increasingly popular and pivotal solution that facilitates the conversion of algorithms expressed in high-level languages (HLLs) such as C/C++/SystemC into hardware descriptions \cite{lahti2018we}. However, HLS as a multi-objective optimization problem (MOOP) necessitates expertise and experience to navigate the optimization among the conflicting objectives such as latency and resource consumption \cite{schafer2019high,liao2023efficient}. Consequently, there is a high demand for the automation of the design space exploration (DSE) process to address the complex tuning challenges for pragmas (or directives).

Heuristic methods offer rapid implementations to yield improved designs \cite{schafer2015probabilistic}. However, their effectiveness is limited by the lack of a thorough understanding of the behaviors of HLS tools. We need to understand how operations are scheduled in order to estimate design performance, but scheduling information is only available after the HLS process \cite{kurra2007impact}. Recently, graph neural networks (GNNs) have been employed for code analysis to predict synthesis results, demonstrating advantages over the preceding DSE methods \cite{wu2020comprehensive, sohrabizadeh2023robust, nobre2016graph, ferretti2022graph, goswami2023machine}. Compilers transform algorithms in HLL into platform-independent intermediate representations (IRs) that encapsulate semantic information \cite{lattner2004llvm, chevalier2007structural, cytron1989efficient}. They are subsequently transformed into control and data flow graphs (CDFGs) for GNN training \cite{cummins2021programl, liang2018flexcl}. However, three critical issues hinder the effectiveness of these methods:

\noindent $1$) The exponential increase in design space size and the time-consuming nature of HLS pose challenges in acquiring sufficient synthesis data for training high-generalization models \cite{ferretti2022graph}.

\noindent $2$) Current methods are tied to specific IR compilers. Using these methods in a different toolchain introduces representation discrepancies. For example, GNNs adopting LLVM IR \cite{lattner2004llvm} does not match HLS tools operating MLIR \cite{ye2021scalehls}. Besides, even when the compilers match, changes in compiler versions require additional workload on the methods. These discrepancies lead to ineffective graph feature extraction \cite{schafer2019high}.

\noindent $3$) The IRs do not adequately contain HLS information, which prevents them from effectively representing design variations \cite{schafer2019high}. As a result, the models leveraging IRs may not fully exploit the design space, leading to suboptimal designs.

Gradient-based inverse design has been widely adopted to retrieve the nanophotonic device designs with desired quality of results (QoR) from large design spaces \cite{gershnabel2022reparameterization}. However, the inherent discrete nature of HLS design spaces keeps from applying the inverse design to DSE for HLS. Although several recent approaches approximated discrete spaces with continuous spaces \cite{maddison2016concrete}, they are not feasible for the inverse design in HLS due to the fact that the pragma effect on the QoR is non-monotonic \cite{sohrabizadeh2023robust}.

To address these challenges, we propose a novel method called \textbf{\textit{Deep Inverse Design for HLS (DID$4$HLS)}}. \textit{DID$4$HLS} focuses on choosing the optimal pragma configurations of the compute-intensive algorithms to achieve designs that are close to Pareto points. We construct the design space with the pragmas of pipelining, loop unrolling, array partitioning, and function inlining. The uniqueness of \textit{DID$4$HLS} lies in its strategy to work on a distribution that samples pragma configurations, with constraints that prevent redundant design samples according to the strategies of HLS tools, such as interactions between pipelining and loop unrolling. Initially, \textit{DID$4$HLS} uses a random pragma distribution to explore a wide range of design possibilities. It then iteratively synthesizes batches of designs generated from the updated pragma distributions, resulting in the IRs that are generally represented in CDFG and crucial in deriving their final hardware designs \cite{kuang2023hgbo}. In this work, we call these IRs as post-HLS IRs (PIRs). These PIRs, along with data on their objectives, are collected into a dataset which is used to train the objective predictors based on Graph Attention Network v$2$ (GATv$2$) \cite{brody2021attentive}. Furthermore, \textit{DID$4$HLS} adopts Variational Autoencoder (VAE) to learn the distribution of design features that determine design performance. They are extracted from the PIRs by the GATv$2$ models. This training is conditioned on the pragma distributions. Through this process, \textit{DID$4$HLS} effectively converges the pragma distribution to approximate Pareto front \cite{schafer2019high}.

In summary, our work addresses the three aforementioned limitations of the prior works as follows:

\noindent $1$) By realizing inverse design, \textit{DID$4$HLS} requires a modest synthesis budget — the limitation of the number of synthesized designs throughout the optimization process — to achieve well-optimized designs, thereby significantly speeding up the synthesis process.

\noindent $2$) Compared to the methods with code analysis, \textit{DID$4$HLS} operates PIRs and avoids representation discrepancies due to reliance on the IR compilers different from the ones adopted by the targeted HLS platforms.

\noindent $3$) In contrast to IRs, PIRs encapsulate essential hardware descriptions and scheduling information, reflecting design variations, leading to an effective DSE process.

We evaluated \textit{DID$4$HLS} through rigorous experiments and quantified its performance in comparison to several prior works. Our results revealed significant improvements in the quality of the Pareto fronts generated by \textit{DID$4$HLS}. In comparison to the best-performing DSE baselines, \textit{DID$4$HLS} improves the average distance to the reference set (ADRS) by $42.8\% $ on average for six benchmarks adapted from Polybench \cite{abella2021polybench} and MachSuite \cite{reagen2014machsuite}. That is, \textit{DID$4$HLS} significantly outperforms the state-of-the-art in achieving a close Pareto front to the optimal. The comparative experiments also demonstrate the high robustness and efficiency of our method. Furthermore, to validate the construction of \textit{DID$4$HLS}, we also conducted an ablation study to explore the benefits of using PIRs and VAE. Results show that our approach achieved an average improvement of $43.0\% $ over the best-performing ablation configurations.
\section{Background}
\subsection{Post-HLS Intermediate Representation}
\begin{figure}
\centering
\includegraphics[width=.45\textwidth]{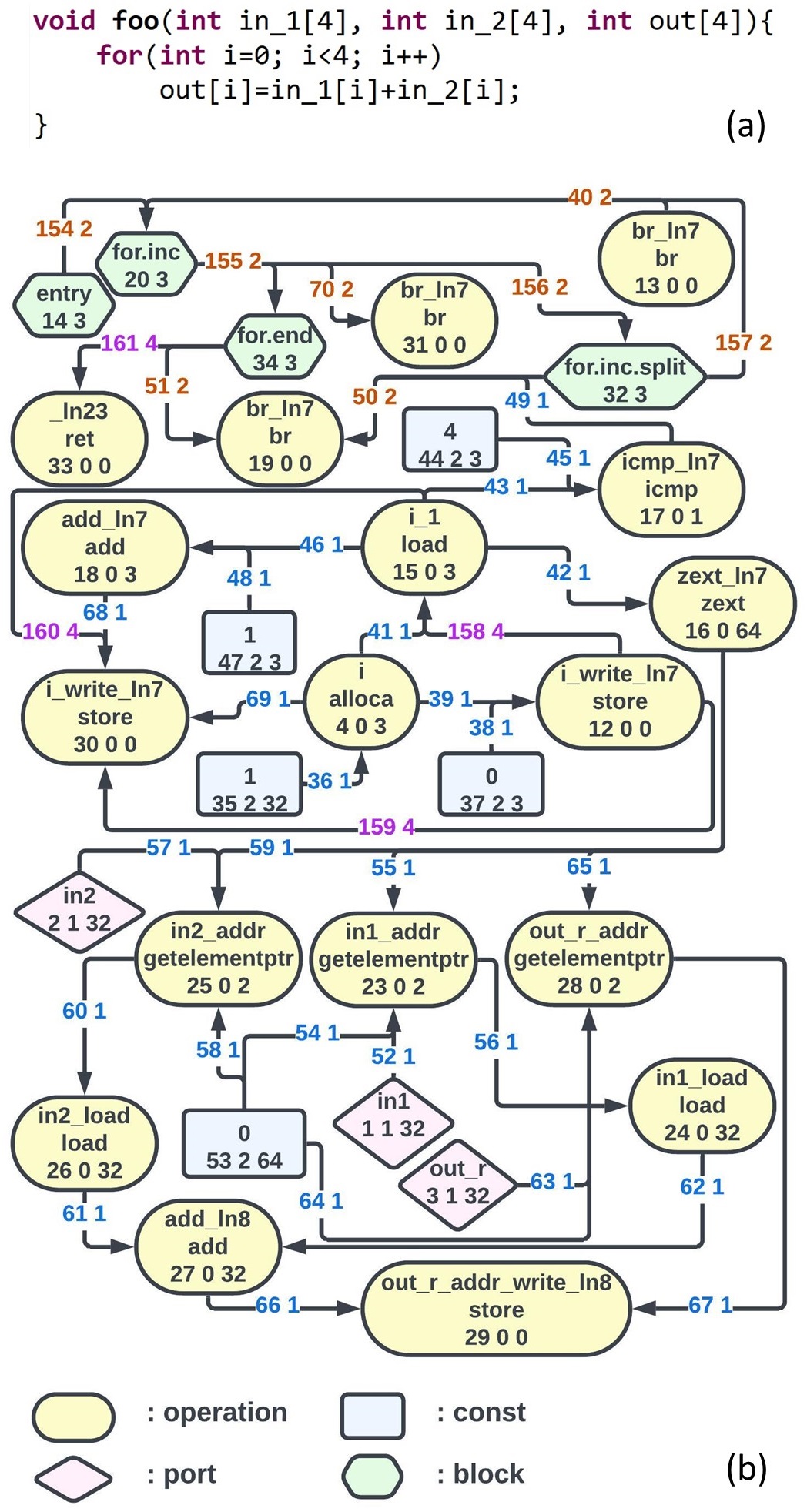}
\caption{(a) A toylike code example. (b) PIR of the code.}
\label{fig:pir}
\end{figure}

The initial HLS phase parses algorithms in HLL into virtual instruction sets called intermediate representations. Subsequent phases entail specifying the order and timing of these instructions under predefined constraints to generate PIRs \cite{kuang2023hgbo}. PIR is a comprehensive representation of the synthesized design in the CDFG format and accurately describes the hardware designs. Consequently, we use PIR to capture hardware design variations induced by pragmas. PIR CDFG is composed of nodes and edges, each labeled with specific types and attributes. A toylike code is illustrated in Fig. \ref{fig:pir} (a) with no pragma applied, and its corresponding PIR generated through Vitis HLS \cite{kathail2020xilinx} is illustrated in Fig. \ref{fig:pir} (b). Node attributes include node types, bitwidths, and consts, while edge attributes include edge types and data flow directions. For example, the node \textit{add\_ln7} of type $0$, indexed at $18$, denotes an \textit{add} operation with a bitwidth of $3$. The edge indexed at $40$ is of type $2$.

\subsection{Graph Attention Networks}
With the aid of the attention mechanism \cite{vaswani2017attention}, Graph Attention Networks (GATs) update node features by learning weighted interactions among nodes and edges \cite{velivckovic2017graph}:
\begin{equation}
\label{eq:static}
e(\mathbf{h}_i,\mathbf{h}_j)=LeakyReLU(\mathbf{a}^\top [\mathbf{W} \mathbf{h}_i \Vert \mathbf{W} \mathbf{h}_j])
\end{equation}

\noindent where $\mathbf{a}$ and $\mathbf{W}$ are trainable parameters, $\mathbf{h}$ denotes node features, $i$ denotes the query node index, $j$ denotes the key node index, and $\Vert $ represents concatenation. Bold font represents matrix or vector in this work. Subsequently, aggregation of all node features gives graph features that are crucial for graph-based DSE methods \cite{sohrabizadeh2023robust}. However, Brody et al. highlighted a significant drawback in GAT, termed static attention \cite{brody2021attentive}, wherein attention rankings of queries collapse toward the dominant key node. This phenomenon arises due to the calculation of attention weights by a single-layer perceptron, as outlined in Eq. \ref{eq:static}, where both linear transformation and activation maintain monotonicity, keeping the rankings of the keys. Despite the introduction of multi-heads in GAT to introduce limited variation in attention rankings, the attention mechanism still experiences a loss of expressiveness. To address this issue, GATv$2$ proposes the substitution of the operation $\mathbf{a}$ as follows \cite{brody2021attentive}:
\begin{equation}
e(\mathbf{h}_i,\mathbf{h}_j)=\mathbf{a}^\top LeakyReLU([\mathbf{W} \mathbf{h}_i \Vert \mathbf{W} \mathbf{h}_j])
\end{equation}

\noindent thereby transforming the single-layer perceptron into an MLP. This modification allows for the approximation of a function capable of selecting any mappings among nodes \cite{hornik1991approximation}, resulting in dynamic attention. Consequently, we employ GATv$2$ to extract graph features in our work.

\subsection{Variational Autoencoder}
Variational Autoencoder (VAE) is an encoder-decoder methodology \cite{doersch2016tutorial} for variational inference \cite{blei2017variational}, aimed at generating data within the distribution of input data $\mathbf{x}$. The encoder models the posterior $q(\mathbf{z}|\mathbf{x})$. The decoder models the likelihood $p(\mathbf{x}|\mathbf{z})$ and maps prior samples into the data within the targeting distribution. Its training objective is expressed below:
\begin{equation}
\label{eq:goal}
\begin{aligned}
\argmin_{\mu ,m,v}\{ &E_{q(\mathbf{z}|\mathbf{x})}[\frac{\Vert \mathbf{x}-\mu (\mathbf{z}) \Vert _2^2}{2c}]\\
&-\sum _i \frac{1}{2}[1+logv_i(\mathbf{x})-v_i(\mathbf{x})-m_i^2(\mathbf{x})] \}
\end{aligned}
\end{equation}

\noindent where the function $\mu $ maps prior sample into the mean of $p(\mathbf{x}|\mathbf{z})$ at the bottleneck. Hyperparameter $c$ governs the variance of $p(\mathbf{x}|\mathbf{z})$, while the functions $m_i$ and $v_i$ calculate the mean and variance, respectively, of the $i$-$th$ dimension of $\mathbf{z}\sim q(\mathbf{z}|\mathbf{x})$. By combining conditioning variables into latent samples, VAE learns conditional distributions.
\section{Problem Definition}
Because HLS optimization is a MOOP, a single distribution is unable to cover the whole Pareto front. For instance, in the context of the trade-off between latency and resource consumption, the distribution prioritizing low latency might be significantly different from the one prioritizing low resource consumption. In order to approximate Pareto front, we pre-define a set of optimization weights $\bm{\lambda }$ sweeping from $0$ to $1$ and we denote the $k$-$th$ weight as $\bm{\lambda }_k$.

We aim to find a design $d$ with the minimal cost for each weight defined as: 
\begin{equation}
\bm{c}_k=\bm{\lambda }_kl+(1-\bm{\lambda }_k)r
\end{equation}

\noindent where $l$ and $r$ denote the standardized latency and resource consumption of $d$. The mean and standard deviation of latency and resource consumption for their standardization are obtained from the designs that are high-level synthesized during the experiments.

We create a vector set $\bm{\Theta }$ in which the vector $\bm{\Theta }_k$ corresponds to $\bm{\lambda }_k$, as a pragma distribution from which pragma configurations are sampled. Each configuration combines with the algorithm in HLL into a design $d$ from a pre-defined design space $S$. Our goal is to identify $\bm{\Theta }_k$ that yields designs close to the optimal design:
\begin{equation}
\bm{d}^*_k=\operatorname*{min}\limits _{\bm{c}_k}d\in S
\end{equation}

Considering the vast size of $S$ and the time-consuming nature of HLS, we adopt an iterative process in which $d$ is sampled with a lower cost mean $E[\bm{c}_k|\bm{\Theta }_k]$ to approach $\bm{d}^*_k$. Therefore, we need to realize an algorithm $A$ that updates $\bm{\Theta }_k$ in iterations:
\begin{equation}
\label{eq:problem}
\begin{aligned}
&E[\bm{c}_k|\bm{\Theta }_k^{i+1}]<E[\bm{c}_k|\bm{\Theta }_k^i]\\
&\bm{\Theta }_k^{i+1}=A(\underset{j}{\cup } \{ [l,r,\bm{\Theta }_k]_d\} _j)\\
&i\in \mathbb{N}\\
&j=1,2,...,i
\end{aligned}
\end{equation}
\section{Methodology}
\subsection{Design Space Interface}
\label{sec:interface}
\begin{figure}
\centering
\includegraphics[width=.45\textwidth]{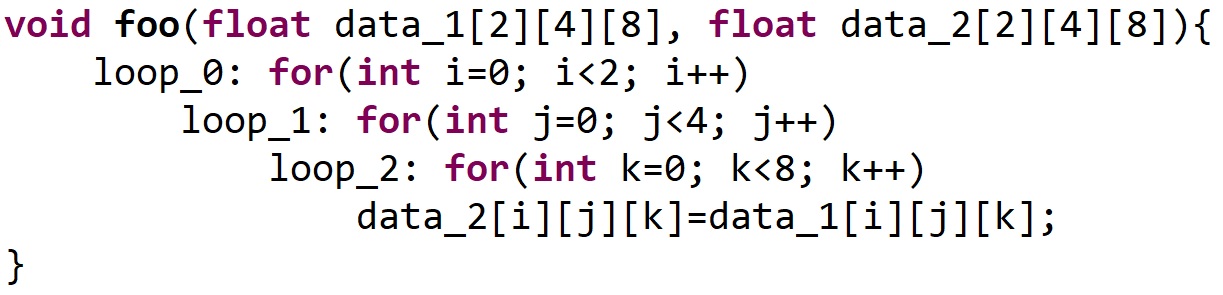}
\caption{An example code with loop labels for HLS.}
\label{fig:loop}
\end{figure}

To start the optimization process targeting pipelining, loop unrolling, array partitioning, and function inlining, an interface is established between algorithms in C++ and our method. Loops within the C++ code files are labeled. The arrays, denoted by $\mathbf{arr}$, introduced below are user-defined that determine the design space $S$. $\mathbf{arr}_{nest}$ describes how the loops nest. An example code with loop labels for HLS is shown in Fig. \ref{fig:loop}. $\mathbf{arr}_{nest}=[-1,0,1]$ means the structure of the three loops. Here, $\mathbf{arr}_{nest}[2]=1$, $\mathbf{arr}_{nest}[1]=0$, and $\mathbf{arr}_{nest}[0]=-1$ mean that the loop of index $2$ is nested under the loop of index $1$, and both are nested under the loop of index $0$. $\mathbf{arr}_{unroll}$ contains user-defined loop unrolling factors. For example, $\mathbf{arr}_{unroll}=[[1,2],[1,2,4],[1,2,4,8]]$ means that $\mathbf{arr}_{unroll}[0]=[1,2]$ allows the loop of index $0$ to be unrolled with a factor of $1$ or $2$, and the other two subarrays are for the rest loops in order. $\mathbf{arr}_{ii}$ contains initiation intervals for a loop if it is pipelined. $\mathbf{arr}_{interface}$ specifies the arrays, at the interface of the top functions, to be partitioned with the type of block or cyclic. $\mathbf{arr}_{p\_itf}$ contains their array partitioning factors. $\mathbf{arr}_{inline}$ contains probabilities for each function to be inlined. $S$ determines the initialization of $\bm{\Theta }$ which is then mapped to align with the deep models, rendering the method inherently self-adaptive. To avoid excessive workloads from redundant design samples with $\mathbf{arr}_{bound}$ containing the loop bounds, we constrain the design sampling process under three HLS rules \cite{xilinx2023vitis,kathail2020xilinx}:

\noindent $1$) If a loop is pipelined, all of its nested loops will be fully unrolled and not pipelined.

\noindent $2$) Pipelining and fully unrolling are mutually exclusive for the same loop.

\noindent $3$) Loops with variable bounds prevent pipelining the loops of the levels above.

To maximize exploration, the probabilities within $\bm{\Theta }$ to sample the pragma values are initialized to Uniform at the beginning of the \textit{DID$4$HLS} process for fully random sampling, except for $\mathbf{arr}_{pipeline}$ that contains probabilities for each loop to be pipelined. The resource required increases exponentially with the depth of loop nests when unrolling outmost loops. For example, if we unroll loop$\_0$ in Fig. \ref{fig:loop}, all its inner loops will be fully unrolled. This often results in a challenge where HLS tools are unable to complete the synthesis process in limited time. Therefore, we initialize $\mathbf{arr}_{pipeline}$ with low probabilities in general, and the outer loops are unrolled with lower probabilities. After sampling, new designs are synthesized to augment the dataset $D$ in each iteration.

\subsection{DID4HLS}
\begin{figure*}
\centering
\includegraphics[width=0.95\textwidth]{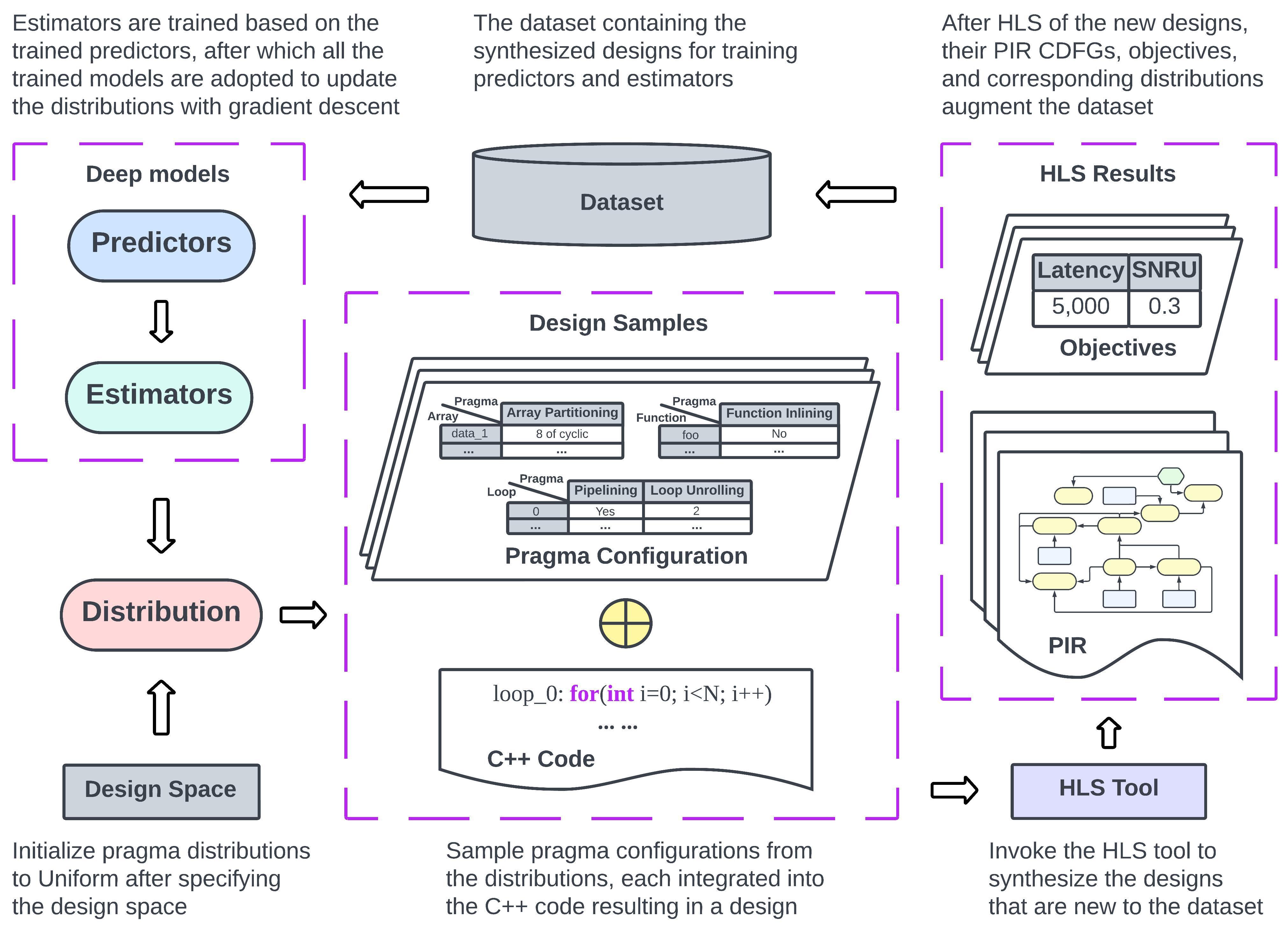}
\caption{Overview of \textit{DID$4$HLS}. After specifying the design space of an algorithm, the corresponding distributions $\bm{\Theta }$ for sampling pragma configurations are initialized to Uniform. In each iteration, the configurations are combined with the algorithm to generate designs described in C++. Invoking HLS tools generates the PIRs and objectives to augment the dataset for training the predictors and estimators, based on which a surrogate is constructed for updating $\bm{\Theta }$, concluding the current iteration.}
\label{fig:overview}
\end{figure*}

We propose \textit{DID$4$HLS} for implementing the algorithm $A$ illustrated in Eq. \ref{eq:problem} to iteratively update $\bm{\Theta }$ in order to sample designs $d$ with minimal costs. The method overview is illustrated in Fig. \ref{fig:overview}.

The pseudocode in Alg. \ref{alg:algorithm} summarizes \textit{DID$4$HLS}. A set of optimization weights $\bm{\lambda}$ sweeping from $0$ to $1$ is pre-defined in order to approximate Pareto front. The arrays introduced in the section \ref{sec:interface} are pre-defined to determine the design space $S$ for each benchmark, and then a set of distributions $\bm{\Theta }$, each $\bm{\Theta }_k$ corresponding to $\bm{\lambda}_k$, are initialized to Uniform (line $3$). $n_1$ pragma configurations are sampled from $\bm{\Theta }$ under the constraint of the three sampling rules. The algorithm in C++ code combining each configuration gives a design $d$. All of them form a set $\{ d\} _{n_1}$ (line $6$). We invoke the HLS tool to synthesize the designs from $\{ d\} _{n_1}$ that are new to $D$ (lines $7$-$8$). The synthesis results include PIR CDFG $g$, design latency $l$ and resource consumption $r$, and the $\bm{\Theta }$ vectors under which the configurations are sampled (line $11$). They are collected into $D$ (line $12$). Then, $l$ and $r$ are standardized (line $13$). Two predictors $P_l(g)$ and $P_r(g)$ are trained with $D$ for predicting standardized $l$ and $r$, respectively (line $14$). The model training takes $g$ as input, standardized $l$ and $r$ as ground truth. Each predictor comprises two parts: an extractor, $F_l(g)$ or $F_r(g)$, for extracting graph feature $\mathbf{f}_l$ or $\mathbf{f}_r$, and an MLP, $MLP_l(\mathbf{f}_l)$ or $MLP_r(\mathbf{f}_r)$, for predicting $l$ or $r$. After training the predictors, $\mathbf{f}_l$ and $\mathbf{f}_r$ of the synthesized designs are extracted (line $15$), combined with corresponding $\bm{\Theta }$ vectors, for training two estimators $V_l$ and $V_r$ that reconstruct $\mathbf{f}_l$ and $\mathbf{f}_r$ conditioned on $\bm{\Theta }_k$ (line $16$). Then, we build a surrogate based on $MLP_l$, $MLP_r$, and the decoders $Dec_l$ of $V_l$ and $Dec_r$ of $V_r$. $n_2$ prior samples $\{ \mathbf{z}_i \} $, combined with $\bm{\Theta }_k$ of current iteration, result in $\hat{\bm{c}}_k$ as the estimation of the cost $\bm{c}_k$ (lines $18$-$21$). Through gradient descent with respect to $\hat{\bm{c}}_k$, $\bm{\Theta }_k$ is updated for several times, where we build a function $\eta $ with the aid of the ADAM optimizer \cite{kingma2014adam} to output updating rates while guaranteeing validity of $\bm{\Theta }_k$ (line $22$). In the next iteration, if the synthesis budget $B$ is reached, an approximate Pareto front will be returned (lines $9$-$10$).
\begin{algorithm}
    \caption{\textit{DID$4$HLS}}
    \begin{algorithmic}[1]
        \State \textbf{Input:} C++ code with loop labels, arrays determining design space $S$, synthesis budget $B$, batch size $n_1$ and $n_2$, optimization weights $\bm{\lambda }$
        \State \textbf{Output:} Approximate Pareto front
        \Statex
        \State Initialize $\bm{\Theta }_k\sim Uniform$
        \State Dataset $D=\varnothing $
        \State \textbf{repeat}
            \State \; \; $n_1$ samples from $\bm{\Theta }:\; \{ d\} _{n_1}$
            \State \; \; $\{ d\} _{new}=\{ d\} _{n_1} \setminus (\{ d\} _D\cap \{ d\} _{n_1})$
            \State \; \; Invoke HLS tool on $\{ d\} _{new}$
            \State \; \; \textbf{if} Number of synthesized designs $==B$:
                \State \; \; \; \; \textbf{return} approximate Pareto front
            \State \; \; $D_{new}=\{ [g,l,r,\{ \bm{\Theta }_k\} ]_{d}\} _{new}$
            \State \; \; $D\leftarrow D\cup D_{new}$
            \State \; \; Standardize $l,r\in D$
            \State \; \; Train $P_l$ and $P_r$ with $g,l,r\in D$
            \State \; \; Extract $\mathbf{f}_l=F_l(g)$ and $\mathbf{f}_r=F_r(g)$ with $g\in D$
            \State \; \; Train $V_l$ and $V_r$ with $\mathbf{f}_l$, $\mathbf{f}_r$, and [$\bm{\Theta }_k]_d \in D$
            \State \; \; \textbf{repeat}
                \State \; \; \; \; $n_2$ samples from $p(\mathbf{z}):\; \{ \mathbf{z}_i\} ,\; i=1,2,...,n_2$
                \State \; \; \; \; $\hat{l}_i=MLP_l(Dec_l(\mathbf{z}_i,\bm{\Theta }_k))$
                \State \; \; \; \; $\hat{r}_i=MLP_r(Dec_r(\mathbf{z}_i,\bm{\Theta }_k))$
                \State \; \; \; \; $\hat{\bm{c}}_{ki}=\bm{\lambda} _k\hat{l}_i+(1-\bm{\lambda} _k)\hat{r}_i$
                \State \; \; \; \; $\bm{\Theta }_k\leftarrow \bm{\Theta }_k-\eta (\bm{\Theta }_k)\sum _{n_2}\partial \hat{\bm{c}}_{ki}/\partial \bm{\Theta }_k$
    \end{algorithmic}
    \label{alg:algorithm}
\end{algorithm}

\subsection{Predictors}
\begin{figure}
\centering
\includegraphics[width=.45\textwidth]{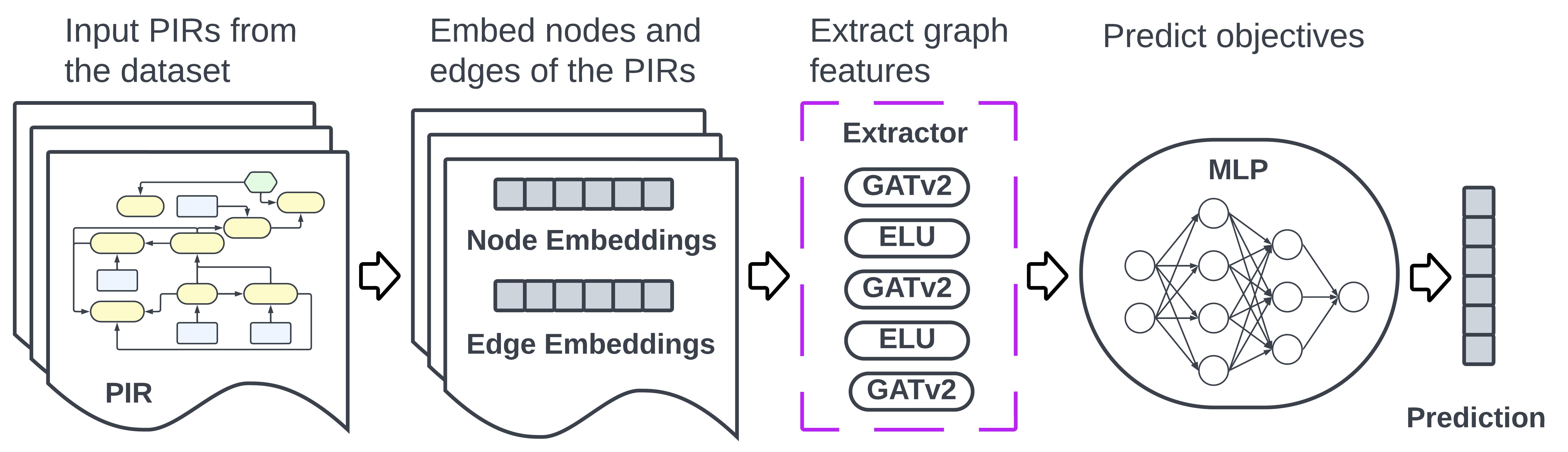}
\caption{Architecture of the predictors. The PIR CDFGs extracted from HLS results are embedded before input into the deep model based on GATv$2$. The model then outputs graph features for the following MLP to predict the objectives.}
\label{fig:predictor}
\end{figure}

Fig. \ref{fig:predictor} illustrates the predictor architecture. We construct $F_l(g)$ and $F_r(g)$, each with three layers of GATv$2$ which demonstrates state-of-the-art performance when applied to graph data \cite{brody2021attentive}. An ELU activation layer is applied between two adjacent GATv$2$ layers \cite{paszke2019pytorch}. Before the PIR data are input into the deep models, their node and edge attributes are embedded.

GATv$2$ updates the node features through the normalized attention weights:
\begin{small}
\begin{equation}
\alpha _{i,j}=\frac{exp[\mathbf{a}^\top LeakyReLU(\mathbf{W}[\mathbf{h}_i \Vert \mathbf{h}_j \Vert \mathbf{e}_{i,j}])]}{\sum _{k\in N(i)\cup \{ i\} }exp[\mathbf{a}^\top LeakyReLU(\mathbf{W}[\mathbf{h}_i \Vert \mathbf{h}_k \Vert \mathbf{e}_{i,k}])]}
\end{equation}
\end{small}

\noindent where $\mathbf{e}_{i,j}$ denotes the edge attribute from node $i$ to $j$, $N(i)$ denotes one-hop neighbor nodes of node $i$. The mean aggregation of features from all nodes results in the graph feature $\mathbf{f}$. Then an MLP follows to predict the objectives. It is important to note that the predictors are trained for feature extraction and not for DSE, since their training data are obtained after HLS.

\subsection{Estimators}
\begin{figure}
\centering
\includegraphics[width=.45\textwidth]{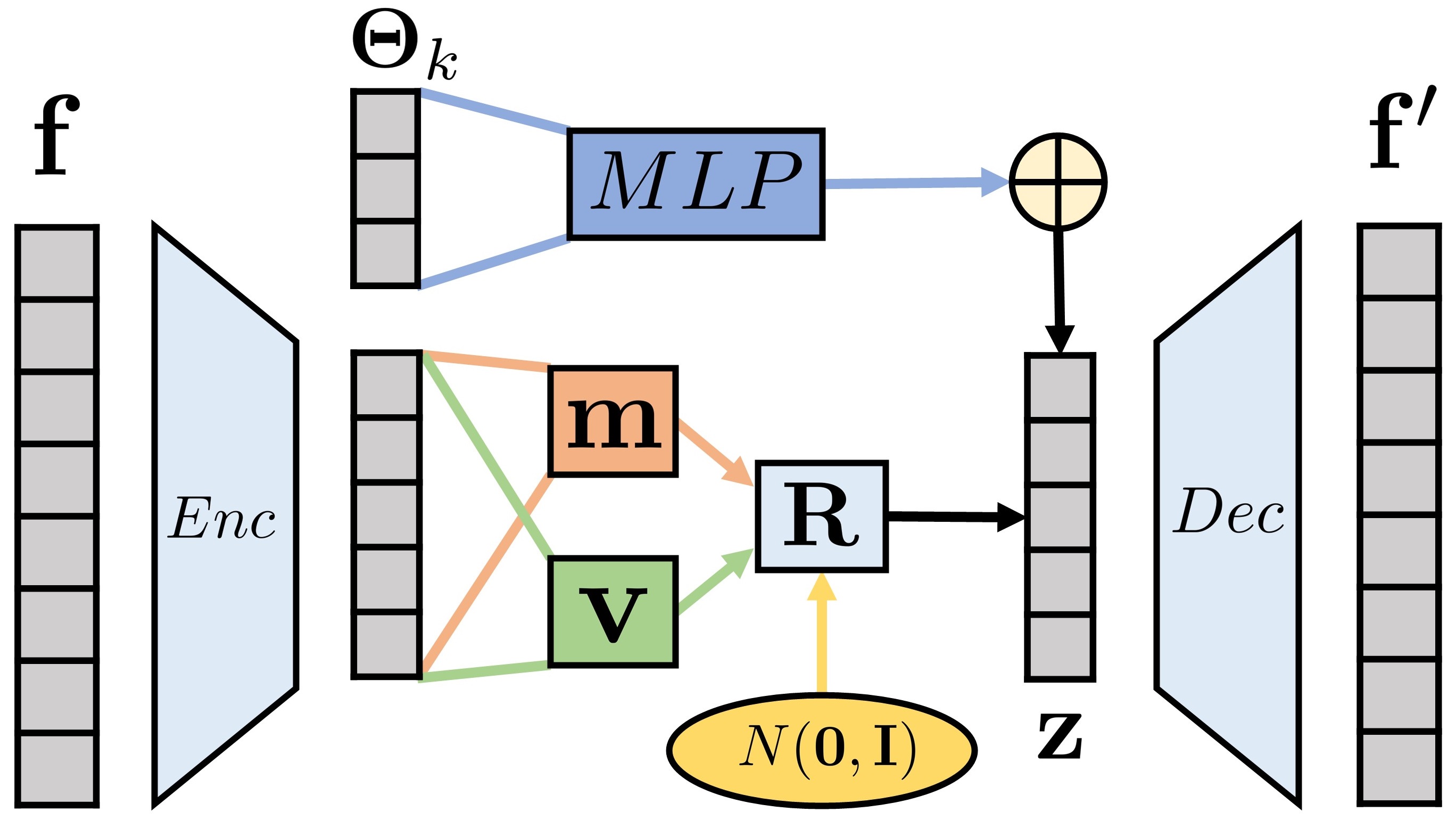}
\caption{Architecture of the estimators based on VAE. $\mathbf{m}$ and $\mathbf{v}$ output the mean and variance respectively of the Gaussian approximation. The node \textbf{R} signifies the reparameterization trick. $Enc$ represents the encoder. $Dec$ represents the decoder tasked with reconstruction of the graph features $\mathbf{f}$, denoted as $\mathbf{f}^\prime $.}
\label{fig:estimator}
\end{figure}

Following the training of $P_l$ and $P_r$, $F_l$ and $F_r$ extract graph features $\mathbf{f}$ of $g\in D$. Then $\mathbf{f}$ and $\bm{\Theta }$ vectors under which the designs are sampled are adopted to train $V_l$ and $V_r$ based on VAE for learning conditional distributions of $\mathbf{f}$. The architecture of these estimators is illustrated in Fig. \ref{fig:estimator}, wherein $Dec$ realizes $\mu $ in Eq. \ref{eq:goal}, $\mathbf{m}$ and $\mathbf{v}$ realize $m_i$ and $v_i$ respectively in Eq. \ref{eq:goal} across all dimensions $i$. We assume the prior distribution as $p(\mathbf{z})=N(\mathbf{0},\mathbf{I})$ \cite{doersch2016tutorial}. After training, the combination of a prior sample $\mathbf{z}\sim p(\mathbf{z})$ and $\bm{\Theta }_k$ transformed by an MLP projector as input into a decoder yields a sample of $\mathbf{f}$ from the estimated distribution.
\section{Experiments}
\subsection{Benchmarks}
Our method emphasizes optimization on algorithms characterized by compute-intensive traits. As such, we evaluated \textit{DID$4$HLS} using six benchmarks adapted from PolyBench and MachSuite:

\noindent $1$) \textit{correlation} has three sets of complex loop nests with intensive computation, introducing diversity in the loop structures. Its loop optimization requires a bandwidth match from memory access arrangement.

\noindent $2$) \textit{covariance} has two sets of complex loop nests with intensive computation. Its complex data access pattern enlarges the room for optimization in parallelization.

\noindent $3$) \textit{gramSchmidt} has one loop nest with five levels performing complicated calculations. The challenges related to both the depth and complexity of the loop structure include addressing data dependencies and managing loop overhead.

\noindent $4$) \textit{aes} is implemented by a flattened loop structure containing nine loops, six of which are wrapped inside functions. This structure increases opportunities for scheduling and resource sharing due to direct data dependencies and frequent function calls.

\noindent $5$) \textit{sort\_radix} heavily depends on memory access patterns. Optimization in loops, arrays, and function calls improves the efficiency of rearranging the data based on digit positions.

\noindent $6$) \textit{stencil\_3d} operates three-dimensional datasets, introducing complex data access patterns. Loop and array optimizations enhance its spatial locality.

The benchmarks and their design spaces tested in our work are accessible on gitHub\footnote{\url{https://github.com/PingChang818/DID4HLS}}.

\subsection{Metric for evaluation}
We adopt ADRS in evaluating the methods in our work \cite{schafer2019high, liao2023efficient}. A design is said to be non-dominated if there is no other design that improves one objective, e.g., latency and resource consumption, without worsening the other, and Pareto front is the set of all such non-dominated designs. ADRS calculates the distance from an approximate Pareto front to the reference Pareto front:
\begin{equation}
ADRS(\Gamma ,\Omega)=(1/|\Gamma|)\Sigma _{\gamma \in \Gamma}\operatorname*{min} \limits _{\omega \in \Omega}f(\gamma,\omega)
\end{equation}

\noindent where $f(\gamma,\omega)=max\{ |\frac{r_\omega -r_\gamma }{r_\gamma }|,|\frac{l_\omega -l_\gamma }{l_\gamma }| \} $, $\Gamma $ is the reference Pareto front, and $\Omega $ is the approximate Pareto front. For each benchmark, $\Omega $ is obtained from the synthesis results of all the tests conducted by one method or ablation configuration, and we obtain $\Gamma $ from all the $\Omega $ datasets. We do not exhaustively search the design spaces because they are too large. For example, that of \textit{stencil\_3d} contains over $10^{15}$ designs. Smaller ADRS denotes closer distance and better performance.

\subsection{Comparison with the baselines}
The two predictors in \textit{DID$4$HLS} share identical architecture, one for latency measured in clock cycles and the other for resource consumption measured in Sum of Normalized Resource Use (SNRU) \cite{ferretti2022graph}:
\begin{equation}
SNRU=\frac{1}{4}\cdot (\frac{FF}{FF^*}+\frac{LUT}{LUT^*}+\frac{DSP}{DSP^*}+\frac{BRAM}{BRAM^*})
\end{equation}

\noindent where $*$ represents available resource. The GATv$2$ layers are implemented using the PyG library \cite{fey2019fast}.

In our method, we set $\bm{\lambda} =[0,0.2,0.4,0.6,0.8,1]$ to approximate the Pareto front. In each iteration, $5$ designs are sampled for each optimization weight and $30$ designs in total are synthesized. $6$ iterations are executed throughout the optimization for each benchmark. The PIR CDFGs are extracted from the $*.adb$ files. We choose four state-of-the-art DSE methods for HLS as the baselines for the comparative analysis. All the baselines share the same design sampling process as \textit{DID$4$HLS} for a fair comparison. We introduce the baselines and provide their implementations below:

\noindent $1$) \textit{GRASP} \cite{schuster2023grasp} is a tree-based heuristic method composed of two steps: Greedy Randomized Construction (GRC) and Local Search (LS). In each iteration, GRC creates a Restricted Candidate List (RCL) containing the most promising pragmas with minimum costs that are estimated by the predictor based on a decision tree. Then the LS phase further refines the pragmas based on the RCL. All the newly synthesized designs update the predictor for the next iteration. We set $100$ estimators for the random forest regressor, cost distance $\alpha $ as $0.7$, and RCL synthesis interval as $8$. In the beginning, we randomly sample $120$ designs and synthesize them to train the predictor. Then we start the GRASP process in iterations until the synthesis budget is reached.

\noindent $2$) \textit{AutoHLS} \cite{rubel2024autohls} constructs predictors with MLP for performance and synthesizability scores of designs. Based on the performance predictor, Bayesian optimization is applied to pragmas and approximates a Pareto front, implemented by Optuna \cite{akiba2019optuna}. In the first iteration, we randomly sample and synthesize $30$ designs to train the predictors. We operate synthesizability scores as the probabilities of designs to be successfully synthesized, and set the theshold as $0.5$. We synthesize $30$ designs from the approximate Pareto front whose scores are predicted over the threshold. After synthesis, we update the models and start the next iteration.

\noindent $3$) \textit{SA-ML} \cite{goswami2023machine} is an annealing-based method. The predictive models are based on XGBoost. We adopt Clang\footnote{\url{https://clang.llvm.org}} to compile the C++ code with pragmas into LLVM IRs. Then, the design features as input to the model are extracted from the IR CDFGs through Programl \cite{cummins2021programl}, including attributes such as the number of nodes of each type, the path length, and the pipelining weight. We set the initial temperature to $10$ with a temperature change rate of $0.1$ and $50$ runs at each temperature. Predictive models are trained with an initial set of $150$ designs randomly sampled and synthesized. Hyperparameter tuning is performed using Comet.ml with Bayesian optimization\footnote{\url{https://www.comet.com}}. We define a set of $30$ optimization weights evenly spaced between $0$ and $1$ to approximate a Pareto front, and randomly sample $30$ designs as the starting point for each weight. Through simulated annealing, each design transitions to its neighbor by randomly selecting one of its pragma and changing to an adjacent value until the stopping condition, which is set to $500$ rejects or the temperature of $10^{-7}$. The candidates are synthesized after the final transitions.

\noindent $4$) \textit{GNN-DSE} \cite{sohrabizadeh2023robust} is an exhaustive search method with a GNN surrogate. We adopt Clang to compile C++ code with pragmas into LLVM IRs. Their CDFGs, obtained through Programl, are input into the GAT-based models for predictions. The Jumping Knowledge Network (JKN) feature identifies the dominant node features of the GAT layers \cite{velivckovic2017graph} and maps them to graph features. Subsequently, an MLP follows to output predictions. We construct two separate models to predict latency and resource consumption. The models are trained with an initial set of $150$ designs randomly sampled and synthesized. In the DSE phase, we randomly sample $10,000$ designs for a comparable runtime and predict their latency and resource consumption with trained models. Then, we approximate a Pareto front and randomly select $30$ designs from the front to synthesize.

\begin{figure*}
\begin{adjustbox}{width=.95\textwidth, center}
\begin{tabular}{ccc}
    \multicolumn{3}{c}{}\\
    \includegraphics[width=50mm]{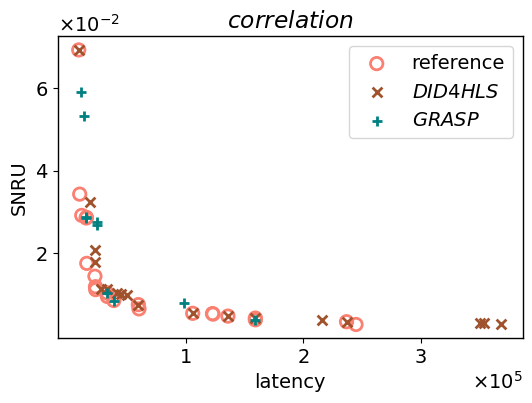}&
    \includegraphics[width=50mm]{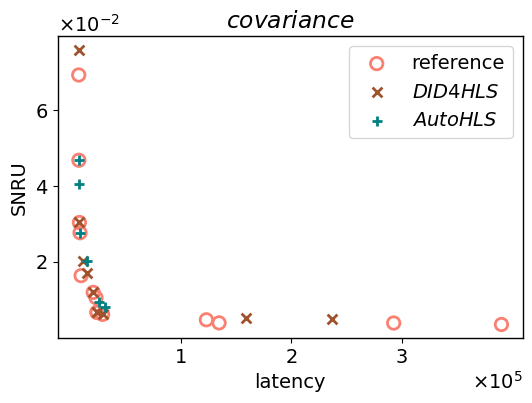}&
    \includegraphics[width=50mm]{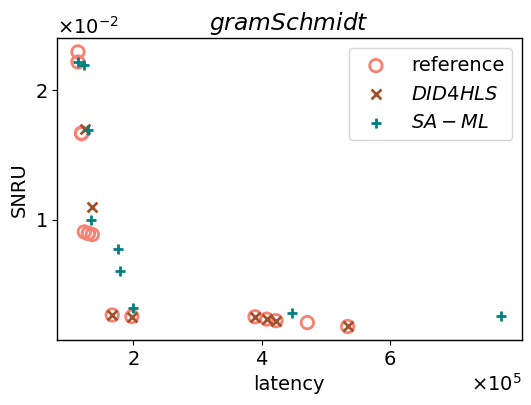}\\
    \multicolumn{3}{c}{}\\
    \includegraphics[width=50mm]{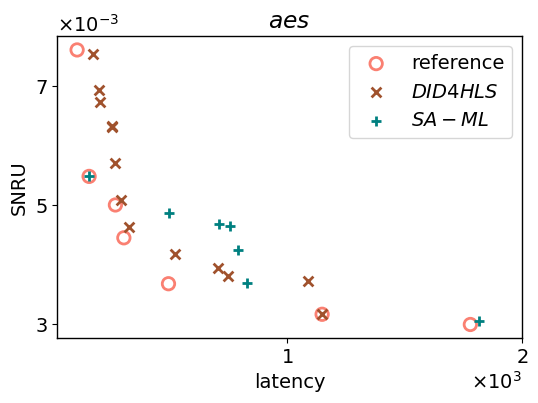}&
    \includegraphics[width=50mm]{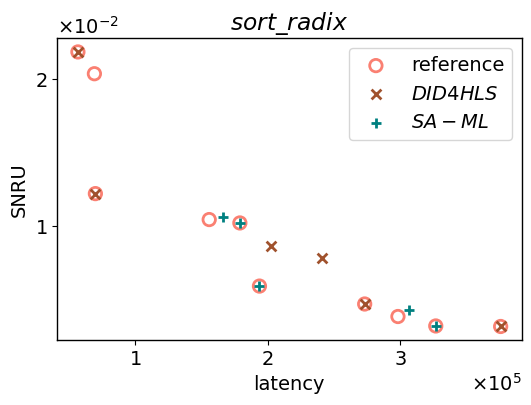}&
    \includegraphics[width=50mm]{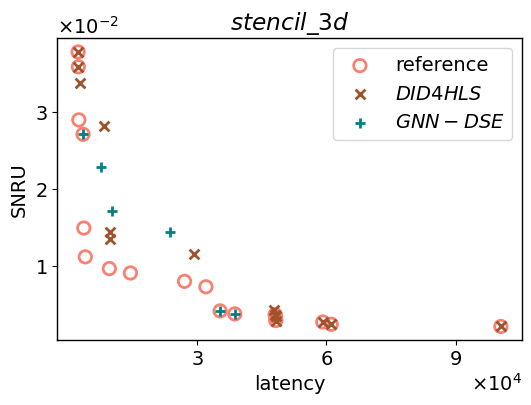}\\
\end{tabular}
\end{adjustbox}
\caption{Comparison between our method and the baselines shows the superior performance of our method, where $\boldsymbol{\circ }$ denotes reference Pareto front, $\boldsymbol{\times }$ and $\boldsymbol{+}$ denote approximate Pareto front of our method and the best-performing baseline respectively.}
\label{fig:front}
\end{figure*}

Tab. \ref{tab:baseline} summarizes the performance comparison between \textit{DID$4$HLS} and the baselines. Our method performed the best on all the benchmarks. The average ADRS improvement over the best-performing baselines achieved by \textit{DID$4$HLS} is $42.8\% $, with up to $74.3\% $ for \textit{sort\_radix}. In the worst case (\textit{aes}), our method improved the ADRS by $23.1\% $. To further demonstrate the effectiveness of our work, Fig. \ref{fig:front} shows the distances of the approximate Pareto fronts by \textit{DID$4$HLS} and the best-performing baselines to the reference Pareto front. Our method offers a wider spread of solutions in latency values for \textit{correlation}, making it more versatile for a range of design requirements. For \textit{aes}, our method significantly outperforms the best-performing baseline, resulting in a distinct gap between the two fronts. Our method achieved more Pareto points than the best-performing baseline for \textit{stencil\_3d} ($8$ vs. $1$), uncovering broader trade-offs and revealing greater potential for diverse designs. Therefore, \textit{DID$4$HLS} demonstrates the superior performance and robustness in improving the design QoR.
\begin{table*}
\caption{ADRS of all methods in the comparative experiments. Bold font highlights the lowest (best) values. Our method performed the best and its improvement over the best-performing baselines are listed.}
\label{tab:baseline}
\begin{center}
\begin{adjustbox}{width=.95\textwidth, center}
\begin{tabular}{ccccccc}
\toprule
Method&\textit{correlation}&\textit{covariance}&\textit{gramSchmidt}&\textit{aes}&\textit{sort\_radix}&\textit{stencil\_3d}\\
\midrule
\textit{GRASP}&$0.262$&$0.481$&$0.359$&$0.515$&$0.701$&$0.612$\\
\textit{AutoHLS}&$0.402$&$0.473$&$0.270$&$0.286$&$1.033$&$0.597$\\
\textit{SA-ML}&$0.780$&$0.766$&$0.170$&$0.286$&$0.514$&$0.487$\\
\textit{GNN-DSE}&$0.356$&$0.486$&$0.330$&$0.356$&$0.610$&$0.436$\\
\textit{DID$4$HLS (ours)}&$\mathbf{0.191}$&$\mathbf{0.166}$&$\mathbf{0.107}$&$\mathbf{0.220}$&$\mathbf{0.132}$&$\mathbf{0.305}$\\
\midrule
Improvement&$27.1\% $&$64.9\% $&$37.1\% $&$23.1\% $&$74.3\% $&$30.0\% $\\
\bottomrule
\end{tabular}
\end{adjustbox}
\end{center}
\end{table*}

\begin{figure}
\centering
\includegraphics[width=.35\textwidth]{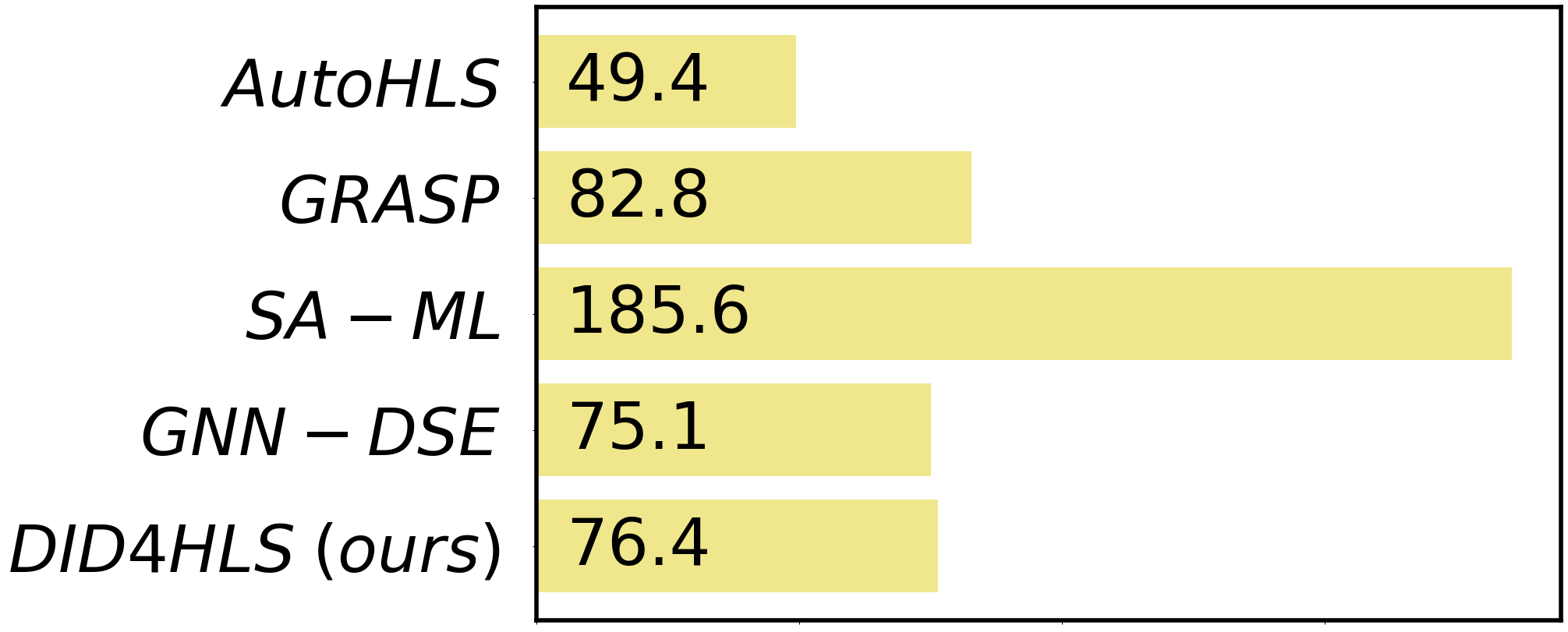}
\caption{The average time in minutes over all the benchmarks consumed by each method.}
\label{fig:speed}
\end{figure}

We also compared the time consumed by each method. Fig. \ref{fig:speed} lists the average time over all the benchmarks. Since designs can take forever to synthesize, we set the synthesis timer to ten minutes, indicating that the design synthesis process will be terminated if it is not completed within this timeframe. \textit{AutoHLS} consumed the least time of $49.4$ minutes. \textit{SA-ML} consumed the most time with $185.6$ minutes. Our method consumed $76.4$ minutes. Although not the fastest, the runtime of our method is justified given its superior performance.

\subsection{Ablation study}
Generative models are particularly relevant for learning conditional distributions, among which VAE, Generative Adversarial Network (GAN), and diffusion model have demonstrated state-of-the-art performance \cite{doersch2016tutorial}. Therefore, we conducted an ablation study to compare their performance.

\begin{figure}
\centering
\includegraphics[width=.45\textwidth]{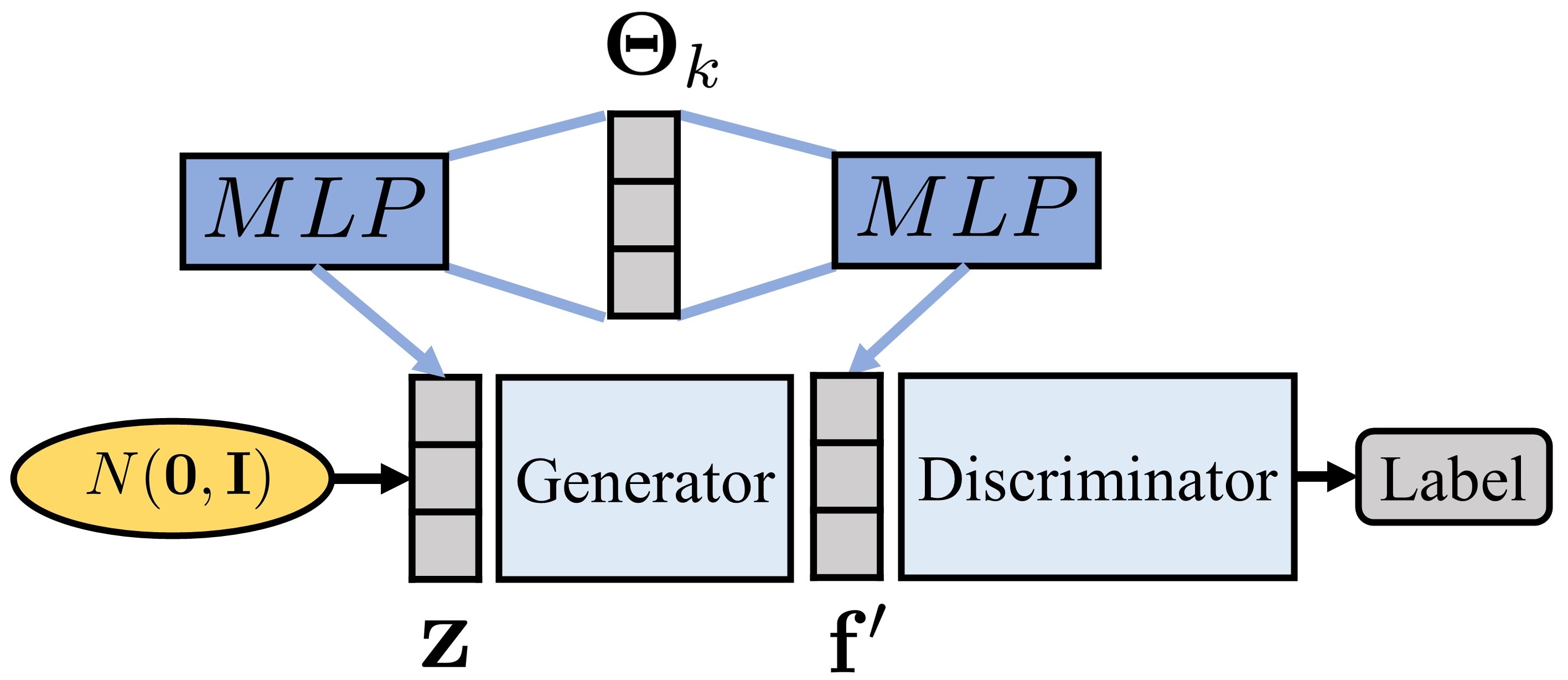}
\caption{Architecture of the estimators based on GAN. The generator and discriminator are constructed with MLP}
\label{fig:gan}
\end{figure}

GAN consists of a generator and a discriminator. The generator takes random noise as input and generates samples similar to the training data. The discriminator is responsible for distinguishing the generated samples from the real ones. As training progresses, both the generator and the discriminator improve in their respective tasks. Fig. \ref{fig:gan} shows the structure of GAN-based estimators, where both the generator and the discriminator have their own MLP projector that transforms $\bm{\Theta }_k$ into a vector concatenated to the input for learning the conditional distribution, $\mathbf{f}^\prime $ denotes the generated graph feature whose ground truth $\mathbf{f}$ is obtained from the predictor's extractor. After training, we adopt the generator to build the surrogate.

\begin{figure}
\centering
\includegraphics[width=.45\textwidth]{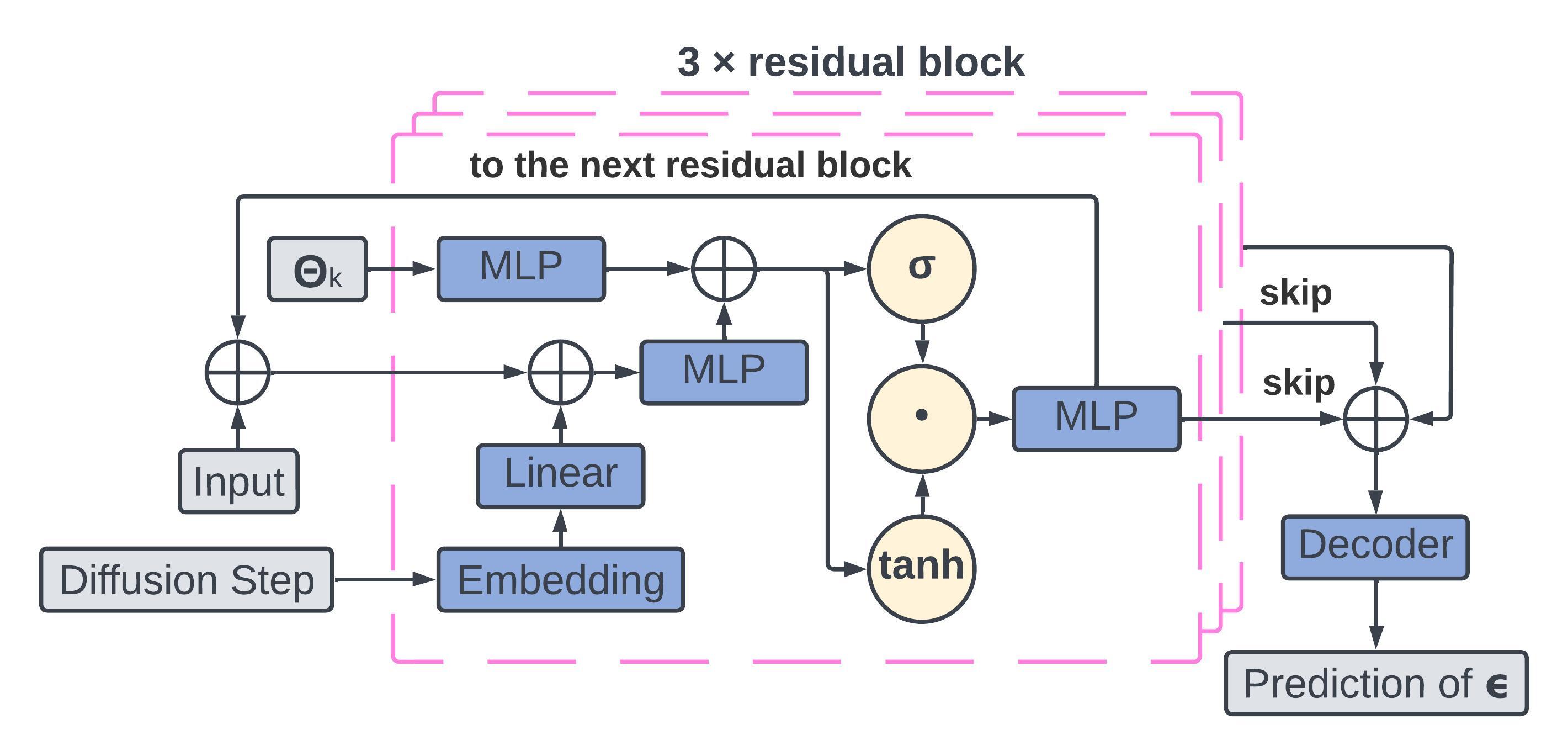}
\caption{Backbone structure of the diffusion model-based estimators. The Gaussian noise $\bm{\epsilon } \sim N(\mathbf{0},\mathbf{I})$.}
\label{fig:diff}
\end{figure}

Diffusion model stems from the idea that a continuous Gaussian diffusion process can be reversed with the same functional form as the forward process. The reverse process converges pure noise to samples in the target distribution. The diffusion process can be approximated with a discrete process given enough diffusion steps. The backbone of diffusion model is responsible for extracting meaningful features from the input data at various diffusion steps to iteratively remove noise. Fig. \ref{fig:diff} shows the backbone structure we adapted from \textit{DiffWave} \cite{kong2020diffwave}, because the gated structure excels at filtering out critical information from the input data. Meanwhile, skip connections mitigate vanishing and exploding gradient problems and capture information at multiple levels of abstraction \cite{kong2020diffwave, chang2024transformer}. We set the number of diffusion steps as $50$ with a quadratic noise schedule from $\beta _1=10^{-4}$ to $\beta _{50}=0.5$ \cite{chang2024transformer}.

To demonstrate the benefits of PIR over LLVM IR in our work, we also conducted a comparative experiment to explore PIR's representational power. In this experiment, we compile the sampled designs into LLVM IR, instead of PIR, to train the predictors, while the estimators are based on VAE.

Tab. \ref{tab:ablation} lists the ADRS values of the ablation configurations, showing the performance to approximate the Pareto fronts achieved by using GAN vs. VAE, diffusion model vs. VAE, and LLVM IR vs. PIR. The complexities of the GAN-based and diffusion model-based estimators are aligned with that of the VAE-based estimators using PyTorch-OpCounter\footnote{\url{https://github.com/Lyken17/pytorch-OpCounter}}, each having around three million parameters. The results show the superior performance of PIR combined with VAE. The average improvement is $43.0\% $ over the best-performing ablation configurations, with up to $71.1\% $ improvement for \textit{sort\_radix}. The least improvement is $19.1\% $ for \textit{aes}.
\begin{table*}
\caption{ADRS of all configurations for the ablation study. Bold font highlights the lowest (best) value. PIR + VAE performs the best and we adopt it in our method.}
\label{tab:ablation}
\begin{center}
\begin{adjustbox}{width=.95\textwidth, center}
\begin{tabular}{ccccccc}
\toprule
Method&\textit{correlation}&\textit{covariance}&\textit{gramSchmidt}&\textit{aes}&\textit{sort\_radix}&\textit{stencil\_3d}\\
\midrule
PIR + GAN&$0.405$&$0.711$&$0.569$&$0.340$&$0.456$&$0.601$\\
PIR + dffusion model&$0.292$&$0.659$&$0.506$&$0.407$&$1.065$&$1.053$\\
LLVM IR + VAE&$0.448$&$0.294$&$0.208$&$0.272$&$0.514$&$0.518$\\
PIR + VAE (ours)&$\mathbf{0.191}$&$\mathbf{0.166}$&$\mathbf{0.107}$&$\mathbf{0.220}$&$\mathbf{0.132}$&$\mathbf{0.305}$\\
\midrule
Improvement&$34.6\% $&$43.5\% $&$48.6\% $&$19.1\% $&$71.1\% $&$41.1\% $\\
\bottomrule
\end{tabular}
\end{adjustbox}
\end{center}
\end{table*}

For each benchmark, we set the synthesis budget to be $180$ solutions. Thus, each method or ablation configuration is allowed to synthesize at most $180$ designs to approximate the Pareto front. We made modifications to the open-sourced code of \textit{GRASP}\footnote{\url{https://github.com/nibst/GRASP_DSE}}, and re-implemented the remaining baselines as closely as possible according to their respective descriptions. All the experiments were implemented using Python on the PyTorch framework \cite{paszke2019pytorch}. The computational hardware utilized included an Intel i$7$-$12700$K processor for HLS and an Nvidia RTX $3090$ graphics card for training deep models. We used Vitis HLS $2023.1$ as the HLS tool. Our simulation platform was set as the Defense-grade Virtex-$7$Q FPGA with part number xq$7$vx$690$t-rf$1930$-$1$I, with a clock constraint of $10ns$. All built-in automatic optimizations were disabled.
\section{Discussion}
The input data for the predictive models in \textit{GRASP} are only pragma configurations. It is hard to iterate toward optimal direction with a limited synthesis budget. Then, designs from LS are easily stuck at local optimum. The surrogate models of \textit{AutoHLS} are hard to capture the complex interactions between pragmas and designs, leading to the inferior performance by the Bayesian optimization. However, the simple input format for the deep models in \textit{AutoHLS}, combined with the high efficiency of Bayesian optimization, results in superior speed performance. \textit{SA-ML} combines simulated annealing and the predictive models for code analysis in LLVM IR, but fails to leverage information from the HLS processes. The sequential nature of its annealing process largely impacts the speed and \textit{SA-ML} consumes the most time. The original implementation of \textit{GNN-DSE} depends on predictive models pre-trained with large datasets. However, in our work, we constrained the training dataset to a small size and observed the overall performance behind \textit{DID$4$HLS}. This highlights the effectiveness of \textit{DID$4$HLS} in capturing HLS behaviors. Our method not only achieves the best QoR, but also operates within a reasonable time frame, demonstrating an optimal trade-off between performance and speed.

GAN, while effective, is known to face challenges such as discriminator overfitting and training stability issues, especially in the cases of limited data availability \cite{karras2020training}. Due to the time-consuming nature of HLS, there is usually a scarcity of training data. The dataset is updated in each iteration and the model must be re-trained thereafter. Thus, the stability issues may accumulate throughout the optimization process and lead to failure in finding the directions along which the distributions converge toward the Pareto front. While diffusion model generates samples with high fidelity and diversity, the stochasticity introduced by multiple diffusion steps hinders the reparameterization trick during the gradient-based update of the distributions. LLVM IR combined with VAE performs far behind \textit{DID$4$HLS}, indicating the limitation of feature extraction from LLVM IR and the importance of using PIR to leverage post-HLS information.

There is still room for improving \textit{DID$4$HLS}. Although it does not depend on a specific compiler to generate IR graphs, transitioning to platforms other than Vitis HLS will necessitate the development of new interfaces parsing the HLS solutions. This issue may hinder the portability of our method. The other issue is that when optimizing larger designs, their PIR CDFGs will be exponentially more complicated. This may be a great burden for our predictors, requiring more computational resources or slowing the optimization. These issues highlight the importance of considering technical limitations in trade-offs between effectiveness and speed when selecting DSE methods for HLS.

To further improve our method, we may also expand the design space accommodating more pragmas such as array reshaping, loop tiling, and data flow. We may also port \textit{DID$4$HLS} to other platforms for additional validation. Implementation of hyperparameter tuning, transfer learning, and online training may be considered to explore more potential.
\section{Conclusion}
In this work, we propose \textit{DID$4$HLS}, a method that learns HLS behaviors, enabling design samples to iterate toward the Pareto front with a low synthesis budget. We demonstrate the effectiveness of \textit{DID$4$HLS} with its improvement on design QoR through the tuning of pragmas for loop pipelining, loop unrolling, array partitioning, and function inlining. The comparative experiments exhibit the superior performance of our method over the four state-of-the-art baselines. It achieved an average improvement of $42.8\% $ on ADRS over the best-performing baselines with high robustness and efficiency. In addition, the ablation study validated the construction of \textit{DID$4$HLS}.

\bibliographystyle{unsrt}
\bibliography{references}

\end{document}